\documentclass[10pt, aps,twocolumn,prl,superscriptaddress,showpacs,floatfix]{revtex4-1}
\usepackage[utf8]{inputenc}
\usepackage[T1]{fontenc}
\usepackage{graphicx}
\usepackage{physics}
\usepackage{amsmath}
\usepackage{amsfonts}
\usepackage{bm}
\usepackage[normalem]{ulem} 
\usepackage{wasysym}
\usepackage{graphicx}
\usepackage{gensymb}
\usepackage{braket}
\usepackage{natbib}
\usepackage{physics}
\usepackage{xcolor}
\usepackage{slashed}

\usepackage{wasysym}

\usepackage[colorlinks=true,citecolor=blue,linkcolor=magenta, breaklinks=true]{hyperref}

\newcommand{\Eq}[1]{Eq.~\ref{#1}}
\newcommand{\Fig}[1]{Fig.~\ref{#1}}
\newcommand{\App}[1]{SM~\ref{#1}}
\def\R{\bm{R}}
\def\bdelta{\bm{\delta}}
\def\M{M}
\def\N{N}
\def\E{\varepsilon}
\def\k{{\bf{k}}}
\def\q{{\bf{q}}}

\usepackage{easyReview}
\def\makeSM{} 
\ifdefined\makeSM
\usepackage{xpatch}
\makeatletter
\patchcmd{\@ssect@ltx}
    {\addcontentsline{toc}{#1}{\protect\numberline{}#8}}
    {}
    {}
    {}
\makeatother
\fi

\def\doctitle{Multiple Dirac Spin-Orbital Liquids in SU(4) Heisenberg Antiferromagnets on the Honeycomb Lattice}
\def\authorA{Manoj Gupta}
\def\authorB{Arijit Haldar}
\def\authorC{Subhro Bhattacharjee}
\def\authorD{Tanusri Saha-Dasgupta}
\def\affiliationI{S. N. Bose National Centre for Basic Sciences, Kolkata 700 098, India}
\def\affiliationII{International Centre for Theoretical Sciences, Tata Institute of Fundamental Research, Bengaluru 560 089, India}

\begin{document}
\title{\doctitle}
\author{\authorA}
\email{gpta.mnj@gmail.com}
\affiliation{\affiliationI}
\author{\authorB}
\email{arijit.haldar@bose.res.in}
 \affiliation{\affiliationI}
 \author{\authorC}
 \email{subhro@icts.res.in}
\affiliation{\affiliationII}
\author{\authorD}
\email{t.sahadasgupta@gmail.com}
 \affiliation{\affiliationI}
\date{\today}

\begin{abstract}
We study the strong coupling model of $d^1$ transition metal tri-halides in the large spin-orbit coupled limit. By considering ab-initio-calculation-inspired hierarchy of hopping pathways of these compounds, SU(4) symmetry is found to emerge at multiple points in the parameter space of the hopping parameters. The resultant Dirac spin-orbital liquids, within the parton mean field description, are distinct. The calculated dynamical structure factor fingerprints this distinctive nature, giving rise to observable effects. This opens up a playground for SU(4) Dirac Spin-Orbital liquid in $d^1$ Honeycomb lattice systems.
\end{abstract}

\maketitle

{\it Introduction :} The interplay of spin-orbit coupling (SOC) and electron-electron correlation can realise a host of novel electronic phases~\cite{witczak2014correlated,Hae-Young,rousochatzakis2024beyond,khomskii2022orbital}. Of particular interest are Mott insulators in the presence of strong SOC, where this interplay naturally leads to frustrated spin Hamiltonians -- the Kitaev model~\cite{KITAEV20062, PhysRevLett.102.017205} and spin-ice~\cite{PhysRevX.1.021002, PhysRevLett.108.067204,rau2019frustrated,RevModPhys.82.53,PhysRevLett.79.2554} systems being prime examples. The SOC-mediated anisotropic interactions can suppress conventional magnetic order and give rise to various quantum paramagnets, including different types of quantum spin liquids (QSLs)~\cite{Balents2010, Savary2017, lee2008end, broholm2020quantum}. Central features of such SOC-magnets are: (1) the emergence of SO locked effective spin $J=1/2$s~\cite{10.1143/PTPS.160.155,PhysRevLett.102.017205,banerjee2016proximate,takagi2019concept,hermanns2018physics,trebst2022kitaev}, and (2) the conspicuous absence of spin-rotation symmetry~\cite{Hae-Young}. 

It was recently proposed that the situation in a class of $d^1$ transition metal tri-halides $\alpha$-MX$_3$~\cite{swaroop1964thermodynamic, swaroop1964electrical, brauer1978handbuch} (M = Ti, Zr, Hf and X = F, Cl, Br) on a honeycomb lattice may extend beyond the two basic tenets mentioned above. Here, the strong SOC leads to effective $J=3/2$ spins. As shown by Yamada {\it et. al.}~\cite{PhysRevLett.121.097201}, starting with a Hubbard model for the case with particular indirect nearest-neighbour (NN) only hopping, an SOC-assisted symmetry-enhanced SU(4) Heisenberg antiferromagnetic Hamiltonian is obtained in the strong coupling limit.

This enhanced symmetry is extremely promising since earlier calculations show that SU(N) magnets, in an appropriate $N\rightarrow\infty$ limit~\cite{PhysRevB.37.3774,PhysRevB.39.11538, READ1989609,PhysRevB.42.4568,PhysRevB.84.174441}, can realize non-trivial QSLs even on bipartite lattices. Indeed, numerical calculations on NN SU(4) antiferromagnetic Heisenberg models~\cite{PhysRevX.2.041013,PhysRevB.107.L180401} on the honeycomb lattice indicate that a U(1) Dirac QSL may be stabilised. Although the QSL proposed by Yamada {\it et.al.} seems to be promising, the indirect hopping-only model of Ref.~\cite{PhysRevLett.121.097201} is at odds with the recent first-principle calculations~\cite{gupta2023abinitio}, which reveal a whole hierarchy of hopping pathways for the underlying $J=3/2$ electrons. The omission of these materials-inspired hopping pathways makes the discovered DSOL and the associated physics rather restrictive. In light of this, it is worth investigating the fate of SU(4) DSOL in $d^1$ honeycomb lattice systems when such realistic hopping pathways are considered.

In this Letter, we arrive at a counterintuitive conclusion that diversity in hopping pathways may not only preserve the stability of DSOL phases, but also add an important parameter space to the problem. We show that this can result in a multitude of distinct Dirac spin-orbital liquids (DSOLs), each characterised by its own unique fractionalized excitations and symmetry implementations. Starting from a generic Hubbard model for $J=3/2$ electrons, we derive the corresponding strong coupling $J=3/2$ spin Hamiltonian. This general model, quite interestingly, hosts several inequivalent realisations of the SU(4) Heisenberg antiferromagnet in the extended parameter space of the hopping pathways. The UV symmetries are found to be implemented differently in the resultant low-energy physics of these inequivalent realisations, thus leading to multiple DSOLs. We show that the origin of this distinction lies in the difference in the pattern of {\it site-dependent rotations} of the effective spins required to bring the corresponding spin Hamiltonian to a manifestly SU(4) invariant form. Therefore, the distinct DSOLs give rise to distinguishable features in dynamical structure factors that are detectable in neutron scattering and/or other spectroscopic experiments. This paves the way to realise several different long-range entangled quantum liquid phases with distinct observable features in effective spin$-3/2$ systems due to (1) emergent symmetries, and (2) non-trivial implementation of the microscopic (UV) symmetries -- engineered via 
strong SOC, hierarchy of hopping pathways and electron-electron correlations.

\begin{figure}
    \centering
    \includegraphics[width=\linewidth]{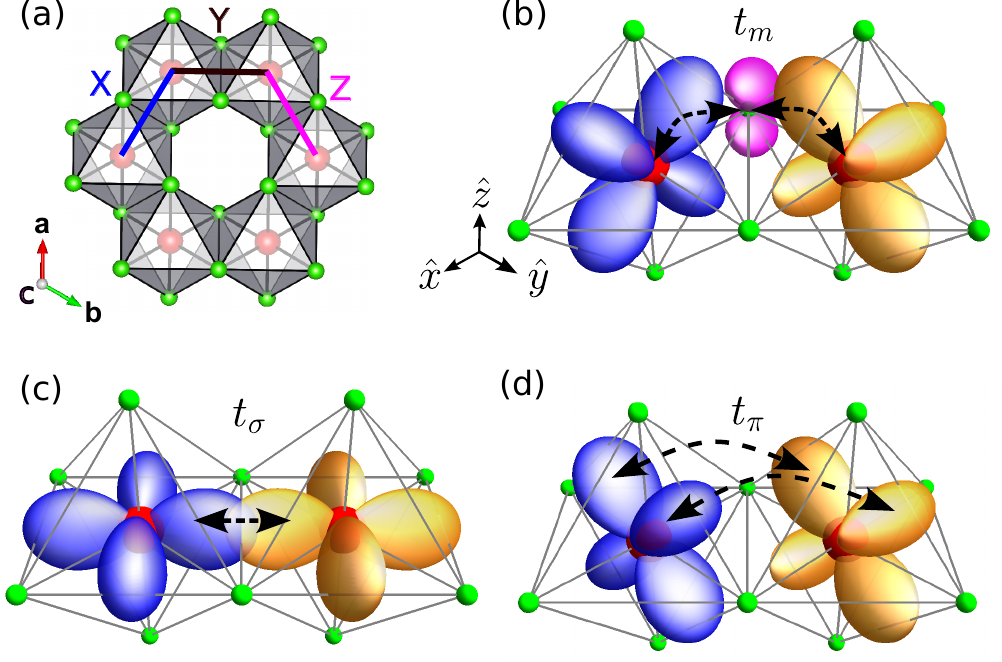}
    \caption{
        Schematic representation of electronic hopping pathways in the honeycomb lattice transition metal tri-halides.
        (a) M-M $X$, $Y$ and $Z$ bonds on the lattice formed by edge-shared $\mathrm{MX}_6$ octahedra. M and X atoms are represented by red and green spheres, respectively.
        (b) Indirect hopping. For clarity, only one ($t_m$) out of two possible indirect hoppings 
        is shown. (c) and (d) Direct $d$--$d$ hopping, with head-on ($t_{\sigma}$) and lateral overlaps ($t_{\pi}$), respectively. 
    }
    \label{fig:pathways}
\end{figure}

\paragraph*{Electronic orbitals, Hopping pathways, SOC and $\&$ SU(4):}  Recent ab-initio work~\cite{gupta2023abinitio} on the honeycomb transition metal tri-halides $\mathrm{MX}_{3}$, comprising of edge-shared MX$_6$ octahedra (Fig. \ref{fig:pathways}(a)) demonstrated that these materials host a rich hierarchy of hopping pathways. In addition to indirect hopping ($t_m$), considered in Ref. \cite{PhysRevLett.121.097201}, where two $t_{2g}$ orbitals at neighboring M sites interact through the intervening, shared $p$ orbital of the halide (cf. Fig. \ref{fig:pathways}(b)); there exists two leading direct hopping pathways, $t_{\sigma}$ and $t_{\pi}$, where the lobes of $t_{2g}$ orbitals overlap, as shown in Fig. \ref{fig:pathways}(c) and Fig. \ref{fig:pathways}(d), respectively. Furthermore, the indirect hopping is ideally represented by two hopping pathways, $t_m$ and $t_{m'}$, only one of them being shown in Fig. \ref{fig:pathways} for clarity. Typically, $t_{\pi}$ $\sim$ - $\frac{t_{\sigma}}{2}$ and  $t_{m}$ $\sim$  -0.05 -- -0.1 $t_{\sigma}$, $t_m$ $>$ $t_{m'}$, setting up a hierarchy of hopping integrals~\cite{gupta2023abinitio}.

In the limit of strong SOC, the $t_{2g}$ orbitals with $d^1$ filling, split into the vacant $J=1/2$ and the active $1/4$-th filled $J=3/2$ orbitals whose physics is captured by  a four orbital Hubbard model~\cite{PhysRevB.108.245106,PhysRevLett.121.097201}
\begin{align}
    H &= \sum_{\langle ij \rangle} \psi_{i}^{\dagger} T_{ij} \psi_{j} + \text{h.c.}+\frac{U}{2}\sum_{i}\psi^{\dagger}_{i}\psi_{i}(\psi^{\dagger}_{i}\psi_{i}-1), \label{eq_j32hubbard}
\end{align}
where $\psi_i (=\psi_{i,1/2}, \psi_{i,-1/2}, \psi_{i, 3/2}, \psi_{i,-3/2})$ describes the $J=3/2$ electrons at each honeycomb lattice site, $i$; $T_{ij}$ are $4\times 4$ hopping matrices on NN bonds $\langle ij\rangle$ incorporating the four hopping pathways --  $t_{\sigma}$,  $t_{\pi}$, $t_{m}$ and $t_{m'}$; and, $U$ is the strength of on-site Hubbard term. The structure of the hopping term, starting with the underlying $t_{2g}$ orbitals, is obtained following the results of Ref. \cite{gupta2023abinitio} and is summarised in the Supplementary Materials (SM)\cite{supp}. 

Central to our analysis is $T_{ij}$, which, on the bond $ij$, can be written as
\begin{align}
T_{ij} = \sum_{\alpha=0}^5 t^{(\alpha)}_{ij}\Sigma^{\alpha}
\label{eq_hopj32}
\end{align}
where $\Sigma^\alpha$ are six (time reversal even) of the fifteen $4\times4$ matrices that generate the four dimensional representation of SU(4)~\cite{supp} and $t^{(\alpha)}_{ij} (\in \mathcal{R})$\cite{note1} are the corresponding amplitudes defined in terms of $t_\sigma, t_\pi, t_m$ and $t_{m'}$.

The symmetry of the Hubbard Hamiltonian in Eq. \ref{eq_j32hubbard} depends on the structure of $T_{ij}$ and can be diagnosed by calculating the directed product of the hopping matrices around a hexagon~\cite{PhysRevLett.121.097201, PhysRevB.108.245106}
\begin{align}
    \prod_{\langle ij \rangle \in \hexagon} T_{ij} =T_{Y}T_{Z}T_{X}T_{Y}T_{Z}T_{X}= W_0\Sigma^0+\sum_{\alpha \neq 0} W_{\alpha} \Sigma^{\alpha}
    \label{eq_loop}
\end{align} 
where $T_\alpha~(\alpha=X, Y, Z)$ are the hopping amplitudes on the three NN bonds (Fig. \ref{fig:pathways}(a)), $W_{\alpha}$ are 16 polynomial functions of the {four leading hopping parameters and $\Sigma^\alpha$ are the corresponding $4\times 4$ SU(4) generators~\cite{supp} and Identity ($\Sigma^0$).


\begin{figure}
     \centering
     \includegraphics[width=1\linewidth]{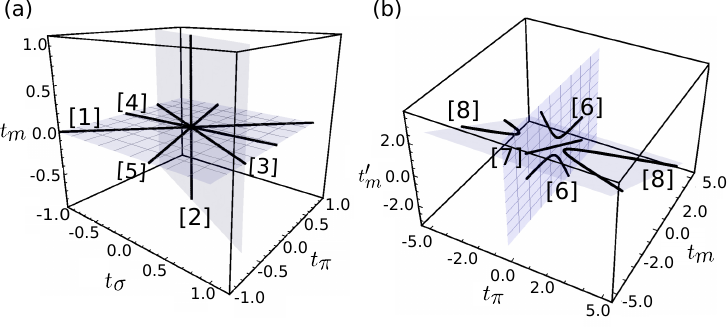}
\caption{The sub-space of hopping pathways with enhanced SU(4) symmetry. (a) In hyper-plane of $t_{m'}=0$: [1] $t_\pi = t_\sigma$, $t_m = 0$;      [2] $t_\pi = t_\sigma = 0$, $t_m \neq 0$;      [3] $t_\pi = -\tfrac{t_\sigma}{2} $, $t_m = 0$;      [4] $t_\pi = \left( \frac{-7 + \sqrt{3}}{5} \right) t_\sigma$, $t_m = 0$;      [5] $t_\pi = \left( \frac{-7 - \sqrt{3}}{5} \right) t_\sigma$, $t_m = 0$.  (b) In hyper-plane of $t_\sigma = 1$:       [6] $t_m \in \mathbb{R}$, $t_\pi = -\tfrac{1}{2}$, $t'_m = -t_m \pm \tfrac{\sqrt{8 t_m^2 + 3}}{2\sqrt{2}}$;       [7] $t_m \in \mathbb{R}$, $t_\pi = -\tfrac{1}{2}$, $t'_m = -\tfrac{t_m}{2}$;      [8] $t_\pi \in \mathbb{R}$, $t_m = \pm \tfrac{1}{3} \sqrt{\tfrac{2}{3}(5 t_\pi^2 + 14 t_\pi - 1)}$, $t'_m = -\tfrac{t_m}{2}$. The planes $t_m = 0$ (shaded light blue), $t_{\pi}=-\frac{t_{\sigma}}{2}$ (shaded gray) and $t'_m = -\tfrac{t_m}{2}$ (shaded gray), $t_{\pi}=-\frac{1}{2}$(shaded light blue) are marked in panels (a) and (b), respectively, as reference.}
     \label{fig:hopping-solutions}
 \end{figure}

In the event the directed loop product of $T_{ij}$s becomes proportional to identity\cite{PhysRevB.108.245106}, implying that the fifteen polynomials $ W_\alpha (\alpha \neq 0)$ vanish leaving only $W_0\neq 0$, an enhanced SU(4) symmetry emerges in the system. This condition ensures that the $J = 3/2$ electron-orbitals do not pick up a direction in the SU(4) space. This requirement was satisfied in Ref. \cite{PhysRevLett.121.097201} by assuming the indirect hopping limit corresponding to  $t_\sigma=t_\pi=t_{m'}=0$ and $t_m=1$. 

The central result of our work is that we find there exists an extended space of hopping parameters with SU(4) symmetry.  This extended parameter space is shown in Fig. \ref{fig:hopping-solutions}(a) and (b), by setting the constraints $t_{m'}$ = 0 and $t_{\sigma} = 1$, respectively. 

The emergent SU(4) symmetry is therefore not restricted to only the particular situation discussed in Ref. \cite{PhysRevLett.121.097201}, rather it appears at multiple distinct sub-manifolds in the entire parameter-space of $t_{\sigma}$, $t_{\pi}$, $t_m$, $t_{m'}$. Among eight possible constraints on $t$'s leading to SU(4) (cf Fig. \ref{fig:hopping-solutions} and the caption), we pick up three specific representative cases in the $t_{m'}=0$ constrained space including the case identified by in Ref. \cite{PhysRevLett.121.097201}, and take them up for the subsequent study. These three scenarios are as follows:

\noindent
\begin{align}
\begin{tabular}{l}
Case-I: \textit{Direct limit}: $t_{\pi} = t_{\sigma} \neq 0,~t_m = 0;~W_{0} = t_{\sigma}^{6}$ \\
\\
Case-II: \textit{Indirect limit}: $t_m \neq 0,~ t_{\sigma} = t_{\pi} = 0;~ W_{0} = -\frac{{t_{m}^{6}}}{27}$\\
\\
Case-III: \textit{$r = -{1}/{2}$ limit}: $t_{\pi} = -\frac{t_{\sigma}}{2},~t_m = 0;~W_{0} = \frac{t_{\sigma}^{6}}{64}$\\
\end{tabular}
\label{eq:cases}
\end{align}
\noindent
Case III introduced above with $r$~=~${t_{\pi}}/{t_{\sigma}} = -1/2$ captures the most realistic situation from a materials perspective~\cite{gupta2023abinitio}.

In these SU(4) symmetric sub-manifolds in the hopping parameter space, the symmetry can be made to manifest via a site-dependent unitary transformation
~\cite{PhysRevLett.121.097201,PhysRevB.108.245106}
\begin{align}
    \psi_i=g_i\phi_i
    \label{eq_gphi}
\end{align}
where $g_i$ are site-dependent $4\times 4$ unitary matrices and $\phi_i$ represent four flavours of fermions that simultaneously diagonalise all the hopping matrices in this  {\it local basis}. The three different cases listed above correspond to three different structures of $g_i$s, as shown in \Fig{fig:sites_presentations}. 

Case-I corresponds to the trivial case where the site-dependent transformations are $g_i=\Sigma^0\equiv\mathbf{1}_4,~~~\forall~i$ such that the SU(4) symmetry is present in the {\it global basis} of $J=3/2$ orbitals, as shown in \Fig{fig:sites_presentations}(a).

\begin{figure}
    \centering
   \includegraphics[width=1\linewidth]{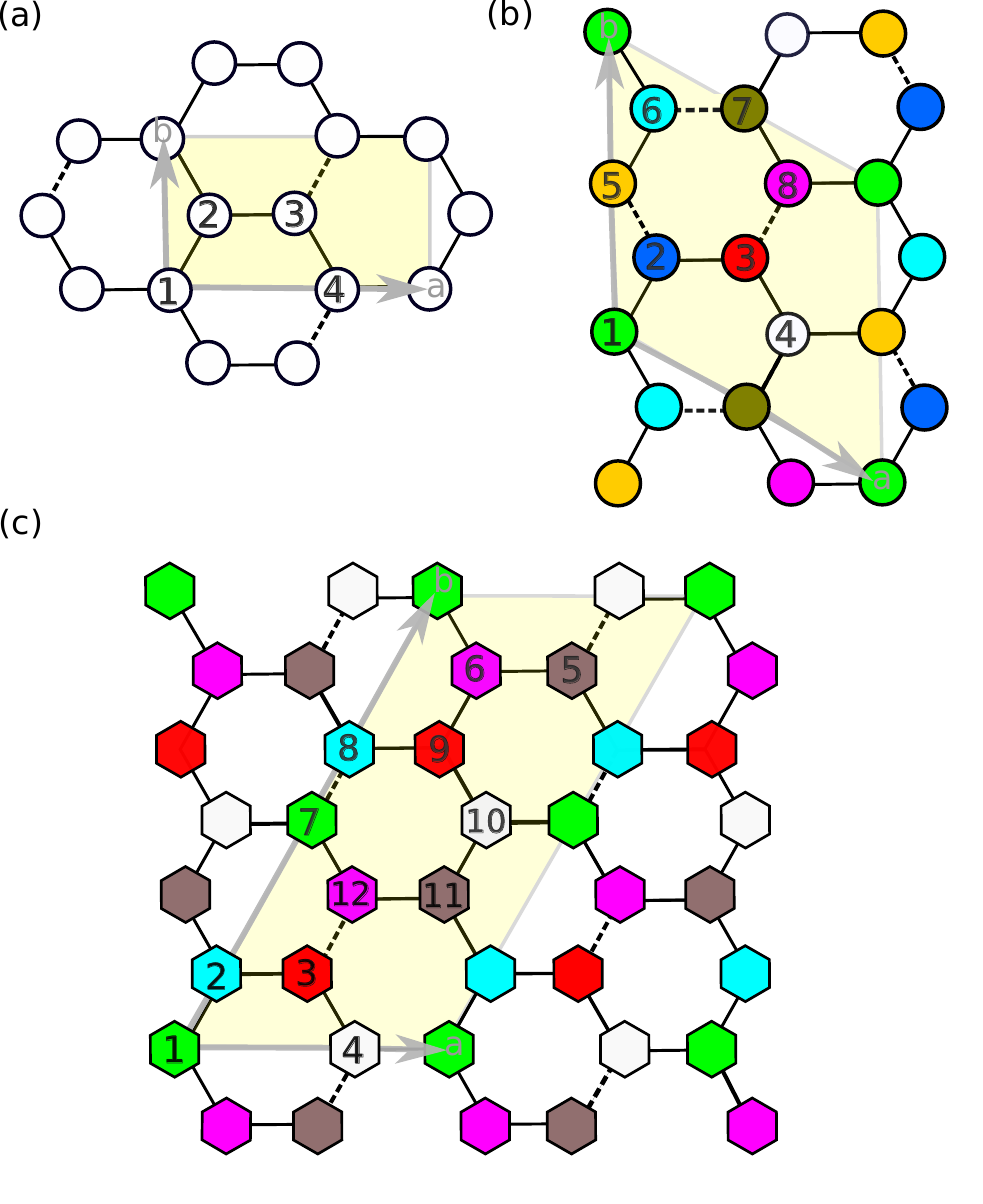}
   \caption{Site-dependent unitary rotation matrices, $g_i$, used for three different cases, discussed in the text. Panels (a) (case-I), (b) (case-II), and (c) (case-III) show the 4-site, 8-site, and 12-site unit cells (indicated by the light yellow shaded boxes), respectively, chosen to be commensurate with the $\pi$-flux pattern. Solid black lines represent hopping with amplitude $+1$, while dotted lines correspond to hopping with amplitude $-1$. In (a) and (b), site-dependent unitary rotations are indicated by distinct colours of spheres, whereas in (c) they are distinguished by colored Hexagons. For the 4-site, 8-site and 12-site unit cells, identity matrix, eight and six different unitary rotations
   are used, respectively.}
\label{fig:sites_presentations}
\end{figure}

Case-II, the limit studied in Ref. \cite{PhysRevLett.121.097201}, has the structure of the local transformations, $g_i$'s, given by Fig. \ref{fig:sites_presentations}(b).  The pattern of the transformations repeats with an eight-site unit cell as shown in the figure.

Case-III corresponds to an unexplored SU(4) representation that provides a description closest to realistic materials. Similar to the first case, here the indirect hoppings are absent, but the direct hoppings are not equal and are close to the ratio observed in the ab initio calculations~\cite{gupta2023abinitio}. Notably, the $g_i$s required are different from case-II, and the pattern for the new $g_i$s requires a 12-site unit-cell to repeat itself, as shown in \Fig{fig:sites_presentations}(c). (See SM~\cite{supp} for technical details). 

Noticeably, for cases-I and III, $W_0>0$ while for case-II, we have $W_0<0$. Hence, the underlying free electron limit of the three cases is quite different, stemming from the fact that in case-II, the electrons (in the local basis) see a $\pi$ flux~\cite{PhysRevB.108.245106} while in the other two cases the flux is absent. In the large $U$-limit of Eq. \ref{eq_j32hubbard}, however, a straightforward second strong coupling expansion for all three cases, in the local basis, leads to a NN SU(4) Heisenberg antiferromagnet
\begin{align}
    \mathcal{H}=\mathcal{J}\sum_{\langle ij\rangle}\sum_{a=1}^{15} O_i^{(a)}O_j^{(a)}
    \label{eq_su4afm}
\end{align}
where $\mathcal{J}>0$ is the antiferromagnetic exchange and 
\begin{align}
    O^{(a)}_i=\phi^\dagger_i\Sigma^{a}\phi_i
\label{eq_su4op1}
\end{align}
are the onsite SU(4) operators in the local basis. 

\paragraph*{The U(1)-Dirac spin-orbital liquid:} The ground state of the SU(4) Heisenberg antiferromagnmet on various lattices have been studied extensively~\cite{PhysRevX.2.041013,PhysRevB.80.064413,PhysRevB.58.9114,PhysRevLett.82.835,PhysRevLett.107.215301,PhysRevB.70.014428,PhysRevB.68.012408,PhysRevB.98.195113,PhysRevB.104.224436}. A systematic study employing a combination of analytical and numerical methods used in Ref. \cite{PhysRevX.2.041013} for the NN SU(4) antiferromagnet (Eq. \ref{eq_su4afm}) on the honeycomb lattice indicate the absence of spontaneous symmetry broken ground state. Further, the VMC calculations also show that a U(1) Dirac QSL is energetically favoured. This was further confirmed by DMRG studies~\cite{PhysRevB.107.L180401}. While Ref.~\cite{PhysRevB.105.L201115} does indicate an alternate possibility of an interesting gapped topological phase, in this work, we focus on the nature of DSOL in the three cases listed above.

A parton description of this DSOL is obtained~\cite{PhysRevB.65.165113,PhysRevB.108.214407,PhysRevB.93.064434} by considering the 4 dimensional representation of SU(4) in terms of four fermionic 
(electronic chargeless) partons -- in our case, in the local basis, $f_{i\alpha}$,($\alpha=1,2,3,4$) that create four $J=3/2$ states at every site, $i$, as
\begin{align}
    f^\dagger_{i\alpha}|0\rangle=|\alpha_i\rangle~~~{\rm with}~~~~f_{i\alpha}^\dagger f_{i\alpha}=1~~~\forall~i,
    \label{eq_su4const}
\end{align}
where $|0\rangle$ is the fermion vacuum and the second equation above is the on-site one-fermion constraint required for faithful representation of the $J=3/2$ Hilbert space.

The SU(4) operators (Eq. \ref{eq_su4op1}) are given in terms of the partons via an analogous expression by replacing the electrons with the fermionic partons ($\phi_i\rightarrow f_i$) to obtain the mean field Hamiltonian~\cite{supp}
\begin{align}
    H_{MF}&=-\frac{\mathcal{J}}{4}\sum_{\langle ij\rangle}\left(\bar\chi_{ij}f^\dagger_{i\alpha}f_{j\alpha}+{\rm h.c.}\right) \nonumber \\
    &+\frac{\mathcal{J}}{4}\sum_{\langle ij\rangle}|\bar\chi_{ij}|^2-\mu\sum_i f^\dagger_{i\alpha}f_{i\alpha}
    \label{eq:HMF}
\end{align}
where $\bar{\chi}_{ij}=\langle f_{j\alpha}^\dagger f_{i\alpha}\rangle$ is the mean field parameter and the last term arises from the constraint in Eq. \ref{eq_su4const}, implemented on an average, which fixes the parton chemical potential to 1/4th filling~\cite{PhysRevX.2.041013}. The mean field DSOL is obtained by choosing
\begin{align}
    |\bar{\chi}_{ij}|=\chi,~~~~\prod_{\langle ij\rangle\in C}{\rm Sgn}[\chi_{ij}]=-1
    \label{eq:chi}
\end{align}
where $\chi(>0)$ is a constant whose value is determined self-consistently while the sign is chosen to implement the $\pi$ flux in each hexagonal plaquette (see \Fig{fig:sites_presentations}) by the second equation in Eq.~\ref{eq:chi}. Thus, the mean field problem reduces to four copies of graphene in $\pi$-flux at 1/4th filling~\cite{PhysRevB.108.245106,PhysRevLett.121.097201}.

To obtain the parton band-structure, one can choose the smallest magnetic unit-cell containing four sites -- consistent with $\pi$-flux (\Fig{fig:sites_presentations}(a)). This choice is sufficient for the global SU(4) (case-I). However due to the non-trivial pattern of the transformations, $g_i$, for the other two cases, it is convenient to choose an eight and twelve site magnetic unit-cell (Fig. \ref{fig:sites_presentations}(b) and (c)), respectively which subsequently (see following) becomes useful in the calculation of the dynamic dipole structure factor. 

The resultant parton band-structure (see SM~\cite{supp}) has degenerate Dirac points located at a pair of time-reversed partner $k$-points of the Brillouin zone (BZ) of the honeycomb lattice, depending upon the choice of magnetic unit cell. In particular~\cite{supp}, we get a pair of Dirac points that are 8-fold degenerate (including fermion flavour) located at 
$k = \pm(\pi/2\sqrt{3}, -\pi/2)$ and at $k=\{\left( -\pi/2\sqrt{3}, \pi/{2} \right)$, $\left( \pi/2\sqrt{3}, \pi/6 \right)$\} for the 4-site and 12-site unit cells shown in Fig. \ref{fig:sites_presentations}, while,  for the 8-site case, the Dirac points are at the BZ centre, {\it i.e.} at the $\Gamma$-point.

The degeneracy and number of Dirac points arise in a manner consistent with the counting associated with the underlying SU(4) symmetry. The low energy theory of the DSOL in all the above cases is therefore a compact $2$+$1$--dimensional QED$_3$ with global SU(8) symmetry comprising four flavors ($f=1, \cdots, 4)$ of 4-component Dirac fermions, $\xi_f$, described by the action $\mathcal{S}=\int d^2{\bf x}\ dt~\mathcal{L}$,
with the Lagrangian $\mathcal{L}$ given by
\begin{align}
    \mathcal{L}=-i\sum_{f=1}^4i\bar\xi_f\slashed{D}_\mathcal{A}~\xi_f -\frac{1}{4e^2} \mathcal{F}_{\mu\nu}^2,
    \label{eq_su8qed3}
\end{align}
where $\slashed{D}_\mathcal{A}=\gamma_\mu D^\mu_\mathcal{A}=\gamma_\mu(\partial_\mu-i\mathcal{A}_\mu)$ with $\gamma_\mu$ being the generators of Clifford algebra~\cite{PhysRevB.108.245106} and $\mathcal{A}_\mu$ being the emergent dynamic U(1) gauge field -- minimally coupled to the fermionic partons -- $\mathcal{F}_{\mu\nu}=(\partial_\mu \mathcal{A}_\nu-\partial_\nu \mathcal{A}_\mu)$. In principle, the gauge field is compact, and the above action needs to be supplemented with amplitudes for instanton events. However, given the large number of fermion flavours, such instanton events are expected to be irrelevant~\cite{murthy1990action,PhysRevLett.98.227202} in a generic parameter regime where the DSOL is stabilised.

Remarkably, due to the difference in the local transformations, $g_i$ (cf. \Eq{eq_gphi}), the underlying microscopic symmetries are differently implemented in each of the discussed cases. This non-trivial implementation of the microscopic symmetries is directly manifested in the dipole correlation function, as discussed in the following.

\begin{figure*}
    \centering
    \includegraphics[width=0.95\linewidth]{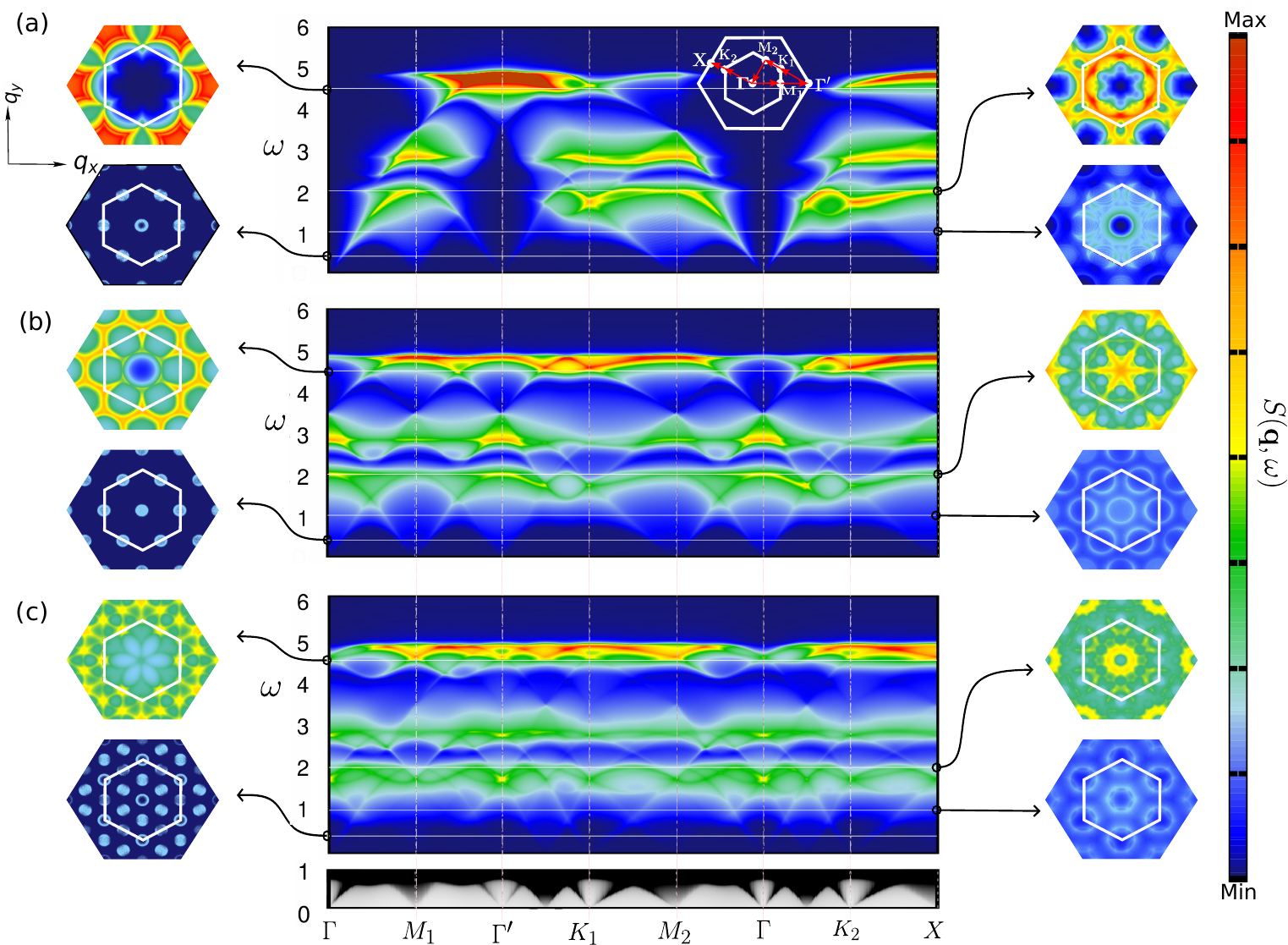}
   \caption{Dynamical dipole structure factor $ S(\mathbf{q}, \omega)$ (Eq. \ref{eq:dyn_struct_fac}) in three cases: Top, middle, and bottom rows correspond to the pure direct limit (case-I) (a), indirect limit (case-II) (b), and \( r = -{1}/{2} \) limit (case-III) (c), respectively. The frequency is measured in units of $\mathcal{J}\chi$. The middle column shows $S(\mathbf{q}, \omega) $ as a function of energy $\omega$ along the high-symmetry path $\Gamma $–$M_1$–$\Gamma'$–$ K_1$-$M_2$-$\Gamma$-$K_2$-$X$ in the first and second BZ of the honeycomb lattice (see inset in top figure). The intensity colour scale is shown on the side. For the \( r = -{1}/{2} \) limit, for better visibility, the low energy part of the spectrum is shown separately in an adjusted contrast of grey scale at the bottom.  Left and right columns show constant $\omega$ cuts of $S(\mathbf{q}, \omega)$ at
   $\omega$ = 0.4, 1, 2 and 4.5. The first BZ is marked by a white line. The colour contrast in constant energy cuts at $\omega$ = 0.4 is adjusted for better visualisation.}
\label{fig:Sq_plots}
\end{figure*}

\paragraph*{Dipole correlation function and spectroscopic signatures of DSOL:} The fifteen SU(4) operators are comprised of three dipole, five quadrupole and seven octupole operators in terms of the underlying $J=3/2$ spins. Of particular interest are the dynamic dipole, ${\bf J}_i=(J^x_i, J^y_i, J_i^z)$, structure factor given by, 
\begin{align}
    \mathcal{S}({\bf q}, \omega)=\frac{1}{N}\sum_{ij}e^{-i{\bf q}\cdot({\bf r_i}-{\bf r_j})}\int dt~e^{i\omega t} \langle {\bf J}_i(t)\cdot {\bf J}_j(0)\rangle
    \label{eq:dyn_struct_fac}
\end{align}
 where (${\bf q}, \omega$) are the momentum and frequency, respectively and $N$ is the number of lattice sites. This expression can be computed by rewriting the dipole operators in terms of the SU(4) generators $O^{(a)}$ (Eq. \ref{eq_su4op1}) followed by decoupling in terms of fermionic partons, which are then calculated, at the mean-field level, via standard many-body techniques~\cite{supp}. 

We present the plots of the dynamical structure factor $S(\mathbf{q}, \omega)$ for the three cases, direct limit (case-I), indirect limit (case-II) and $r=-1/2$ limit (case-III) 
in \Fig{fig:Sq_plots} \cite{footnote2}. The main panels plot $S({\bf q}, \omega)$ along the high-symmetry path in the BZ of the underlying honeycomb lattice. We also show constant $\omega$ cuts across the entire first BZ for four representative values of $\omega$ for all the DSOLs. 

Conspicuous in the constant $\omega$ cuts are three different realisations of the microscopic six-fold symmetry in the three DSOLs, while, in the local basis,  all of them are identical. In particular, while the lowest frequency ($\omega=0.4$) spectral weight for case-I and II appear near the $\Gamma, \Gamma'$ and the $M$ points, case-III has additional spectral weights near the $K$-points as well as midway along the $\Gamma-K-\Gamma'$ line. This difference is further highlighted in the main panel (bottom row), which shows the cut along the high-symmetry directions. Remarkably, for case-I, the circular intensity pattern around the $\Gamma$ point appears to stem from a well-defined {\it quasi-particle} like branch emanating from the $\Gamma$ point, as is evident from the sharp intensity line in the middle panel of the top row. The corresponding branches for cases II and III (middle panel -- middle and bottom rows) are comparatively more diffused.
At $\omega=1.0$, the spectral weight shifts away from the $\Gamma$ and $M(K)$ points for case-II (III) with distinguishable circular patterns, while in case-I, a six-fold flower-petal like structure is seen along the $\Gamma-K$ directions in addition to the central circular minima (maxima) around the $\Gamma (\Gamma')$ point. This serves as a distinguishing feature of case-I from the other two. These features evolve into a six-folded star-like pattern of high intensity for case-II at $\omega=2$, while at the same energy, the intensity is concentrated along the BZ boundary for case-I and zone centre for case-III. Interestingly, the high symmetry cuts in the middle panel reveal that for case II, the higher intensity shifts to the $\Gamma-K$ direction just below $\omega=2.0$. Finally, at the highest frequency ($\omega=5.0$), the intensity is generically concentrated along the boundary of the first BZ and the $\Gamma'$ point in all the cases (with differing details), indicating short-range correlations at this energy scale. 

\begin{figure}
    \centering
   \includegraphics[width=1\linewidth]{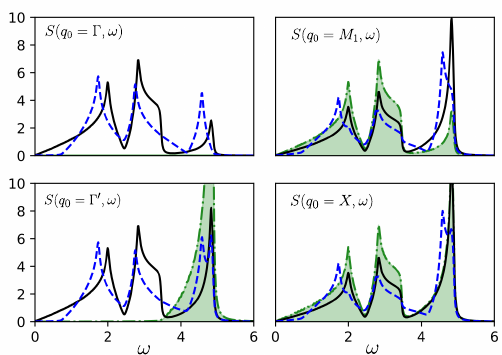}
   \caption{$S(q_0,\omega)$ evaluated at different high-symmetry points in first and second BZ of the honeycomb lattice, $\Gamma$, $M_{1}$ (c), $\Gamma'$, and $X$. The green shaded, black-solid and blue-dotted lines refer to case-I, case-II and case-III, respectively.}
\label{fig:global_probes}
\end{figure}

{The momentum integrated dipole dynamic structure factor $S(\omega)=\sum_{\bf k}S({\bf k},\omega)$ is the same for all three cases which only differ in the momentum distribution of the spectral weights due to the static rotations, $g_i$. Fig. \ref{fig:global_probes} shows plot $S(q_0,\omega)$ for few selected high symmetry momenta $q_0 = \Gamma,\ M_1$, $\Gamma'$ and $X$. 
All the plots show a three-peaked structure, corresponding to three prominent high intensity parts -- 
around $\omega=2.0, 3.0$ and $5.0$. However, the detailed features and weights among
the three peaks vary among the three DSOLs. The most striking distinction is observed 
in $q_0 = \Gamma$ and $\Gamma'$, where the case-I DSOL shows zero response for all energies, $\omega$, at $\Gamma$ and contributes only to the third peak in $\Gamma'$. Further, the case-III DSOL shows a measurable response only above a certain energy threshold, and the third peak in energy (at $q_0 = \Gamma'$, $X$ and $M_1$) for this DSOL phase develops a bifurcated feature unlike the other cases.} 

The natural experimental probe to detect the dipole structure factor is neutron scattering. Two comments to this end -- (a) while in the atomic orbital limit, the Land{\'e} $g$ factor identically vanishes, hybridization with the halide $p$ orbitals gives rise to finite dipole moment when can then couple to the neutrons~\cite{PhysRevB.82.174440,PhysRevB.100.045142,PhysRevLett.118.217202}, and, (b) In addition to the dipole, the octupoles, in principle can also couple to the neutrons. However, such couplings are expected to be weaker. Keeping these in mind, the above distinct signatures of fractionalisation in the three DSOLs can be probed via neutron scattering probing $S({\bf q}, \omega)$ to yield characteristic signatures of such phases. Note that due to the subtle momentum-dependent signatures, the distinction of the three DSOLs is not apparent in $S(\omega)$. However, Raman or infrared scattering probing the ${\bf q}=0$ can distinguish between the three cases, assuming that the signals are dominated by dipolar correlations.

{\it Summary and Outlook :} We show that the SOC-enabled spin-orbit locking~\cite{PhysRevB.108.245106,bhattacharjee2012spin} in $d^1$ transition metal tri-halides can lead to several different U(1) DSOLs. In particular, allowing for possibilities of different hopping pathways, we achieve distinct realisations of microscopic symmetries in three DSOLs with imprints in the experimentally measurable dipole dynamical structure factor. One of these DSOLs (case-III) lies close to the microscopic parameter regimes of these materials according to recent ab-initio calculations~\cite{gupta2023abinitio}.

Our study, thus, opens up a bigger canvas and raises a compelling possibility where tuning of hopping paths via chemistry - choices of different metals, anions, or via straining may result in a variety 
of DSOLs with distinct observable features.  Specifically, this may result in situations (case-II and case-III) distinct from the SU(4) Kugel-Khomskii Hamiltonian~\cite{kugel1982jahn,PhysRevX.2.041013,PhysRevB.80.064413,PhysRevLett.81.3527,PhysRevB.91.155125,PhysRevLett.81.3527}. 

The present mean field theory treatment only accounts for the fractionalised fermionic parton contributions. This invariably gets renormalised by long-wavelength gauge fluctuations (emergent photons), which would further contribute to the dynamic structure factor, particularly near the $\Gamma$ point. However, numerical calculation of Gutzwiller-projected mean field state~\cite{PhysRevB.108.214407} preserves the qualitative features of the mean field result, as expected from the large N-limits of SU(N) models~\cite{READ1989609,PhysRevB.37.3774,PhysRevB.39.11538}. 

A major challenge from a materials perspective is that the MX$_3$ compounds show a tendency to dimerize~\cite{cryst7050121,Trans_TiCl3, troyanov1991x, Trans_TiBr3} which can be suppressed by intercalation and nano-structuring, as found in IrTe$_2$~\cite{PhysRevLett.108.116402, IrTe2_Yoshida}. Such a dimer state as well as others,\cite{PhysRevB.108.245106,chaubey2025superconductivity,PhysRevB.87.224428,seifert2024spin,PhysRevLett.125.257202,van2000plaquette,van2001spontaneous,PhysRevB.93.064434}
can be thought of as instabilities of the DSOL driven by residual short-range parton-parton interactions arising from deformation of the SU(4) Hamiltonian by other material relevant interactions
~\cite{PhysRevB.108.075111,PhysRevB.100.205131,PhysRevB.107.L180401}. 
These, and other relevant instabilities~\cite{PhysRevB.87.224428,seifert2024spin,PhysRevLett.125.257202,van2000plaquette,van2001spontaneous,PhysRevB.93.064434} can be investigated by studying the symmetry of the fermion-bilinears of Eq. \ref{eq_su8qed3}.

{\it Acknowledgements :} We thank Basudeb Mondal, Ankush Chaubey and Vijay B. Shenoy for previous collaboration on the related topic and Karlo Penc for fruitful discussions. M.G. acknowledges CSIR, India, for the senior research fellowship (Grant no. 09/575 (0131) 2020-EMR-I). AH acknowledges support from DST India via the ANRF grant SRG/2023/000118. SB acknowledges funding by the Swarna Jayanti fellowship of SERB-DST (India) Grant No. SB/SJF/2021-22/12 and the Department of Atomic Energy, Government of India, under project no. RTI4001. T.S-D acknowledges J.C.Bose National Fellowship (grant no. JCB/2020/000004) for funding. SB and T.S-D acknowledge funding from DST, Government of India (Nano mission), under Project No. DST/NM/TUE/QM-10/2019 (C)/7. 

\bibliography{REF}

\begin{thebibliography}{73}
\bibitem [{not()}]{note1}%
  \BibitemOpen
  \href@noop {} {}\bibinfo {note} {Though we used the same notations, $t^{(\alpha)}_{ij}$ in Eq. (2) refers to expansion of the hopping matrix in J-3/2 basis as $\Sigma$'s, while $t_{\sigma}$, $t_{\pi}$, $t_{m}$ and $t_{m}'$ denote the hopping pathways in $t_{2g}$ basis.}\BibitemShut {Stop}%
\bibitem [{foo()}]{footnote2}%
  \BibitemOpen
  \href@noop {} {}\bibinfo {note} {Note that these dipole correlations are different from the SU(4) color correlations plotted in Ref-PhysRevX.2.041013.}\BibitemShut {Stop}%
\end{thebibliography}
\ifdefined\makeSM
\appendix
\onecolumngrid
\renewcommand{\appendixname}{}
\renewcommand{\thesection}{{S\arabic{section}}}
\renewcommand{\thesubsection}{{\Alph{subsection}}}
\newpage\clearpage
\renewcommand{\theequation}{\thesection.\arabic{equation}}
\renewcommand{\thefigure}{\thesection.\arabic{figure}}
\setcounter{page}{1}
\setcounter{secnumdepth}{3}
\setcounter{figure}{0}
\setcounter{equation}{0}

\onecolumngrid

\begin{center}
    \Large \textbf{\underline{Supplementary Materials}} \\[0.5em]
    \large \textbf{\doctitle} \\[1em]
    \normalsize
    {\authorA\textsuperscript{1}}, {\authorB\textsuperscript{1}}, {\authorC\textsuperscript{2}}, and {\authorD\textsuperscript{1}} \\
    [1em]
    \textit{\textsuperscript{1}\affiliationI} \\
    \textit{\textsuperscript{2}\affiliationII} \\

    (Dated: \today)
\end{center}
\vspace{1em}

\section{Low energy electronic physics of \texorpdfstring{$\alpha$-}{}MX\texorpdfstring{$_3$}{}}
\label{appen_electronic}

The minimal model capturing the low-energy electronic physics is given by~\cite{gupta2023abinitio}
\begin{align}
    {\cal H} = {\cal H}_t +H_U,
    \label{eq_t2ghubbard}
\end{align}
where 
\begin{align}
    H_U=\frac{\tilde{U}}{2}\sum_i\sum_{\alpha} \sum_{\eta,\eta'}\Psi^\dagger_{i\alpha\eta}\Psi_{i\alpha\eta}(\Psi^\dagger_{i\alpha\eta'}\Psi_{i\alpha\eta'}-1)
\end{align}
where $\tilde{U}$ is the magnitude of the on-site Coulomb repulsion for the electrons at the $i$th site of the lattice in the $t_{2g}$ orbital, $\alpha(=xy, yz, xz)$ with spin $\eta(=\uparrow, \downarrow)$. $\mathcal{H}_t$ is the single-particle hopping Hamiltonian that was derived in Ref. \cite{gupta2023abinitio}. The notations followed are the same as in Ref. \cite{gupta2023abinitio}.

The SOC term in $\mathcal{H}_t$ splits the $t_{2g}$ manifold into the upper $J=1/2$ doublet and lower $J=3/2$ quartet, with the latter forming the active orbitals for the $d^1$ electronic configuration. Thus, in the large SOC limit, the effective hopping Hamiltonian is obtained by projecting to the four $J=3/2$ orbitals, resulting in the Hamiltonian given by Eq. \ref{eq_j32hubbard} in the main text. 

\subsection{Hopping Hamiltonian on X, Y and Z bonds in the basis of $J=3/2$ electron orbitals}

The explicit form of the hopping matrix on the $Z$-bond (see Fig. \ref{fig:pathways}(a) in main text) is obtained by projecting ${\cal H}_t$ in Eq. \ref{eq_t2ghubbard} to the $J=3/2$ manifold and is given by~\cite{gupta2023abinitio}

\begin{align}
    T_{Z} =
\left(
\begin{array}{cccc}
 \frac{1}{3} \left(t_{\pi} + 2 t_{\sigma}\right) & 0 & -\frac{(1+i) t_{m'}}{\sqrt{3}} & \frac{i t_{\text{m}}}{\sqrt{3}} \\
 0 & \frac{1}{3} \left(t_{\pi} + 2 t_{\sigma}\right) & -\frac{i t_{\text{m}}}{\sqrt{3}} & \frac{(1-i) t_{m'}}{\sqrt{3}} \\
 -\frac{(1-i) t_{m'}}{\sqrt{3}} & \frac{i t_{\text{m}}}{\sqrt{3}} & t_{\pi} & 0 \\
 -\frac{i t_{\text{m}}}{\sqrt{3}} & \frac{(1+i) t_{m'}}{\sqrt{3}} & 0 & t_{\pi} \\
\end{array}
\right)
\label{eq:T_z}
\end{align}

The hopping on the X and Y bonds can be obtained via the $C_3$ rotation given by, $U_{C_{3}}$ of the form
\[
-\frac{1}{4} 
\begin{bmatrix}
    1 - i & 1 + i & (1 + i) \sqrt{3} & (1 - i) \sqrt{3} \\
    -1 + i & 1 + i & (1 + i) \sqrt{3} & (-1 + i) \sqrt{3} \\
    (1 - i) \sqrt{3} & (-1 - i) \sqrt{3} & 1 + i & -1 + i \\
    (-1 + i) \sqrt{3} & (-1 - i) \sqrt{3} & 1 + i & 1 - i
\end{bmatrix}
\]

such that $T_{Y}=U_{C_{3}}\cdot T_{X}\cdot U_{C_{3}}^{\dagger}$ and $T_{X}=U_{C_{3}}\cdot T_{Z}\cdot U_{C_{3}}^{\dagger}$. 
The non-zero amplitudes for the hopping on the NN bonds for the $J=3/2$ orbitals (Eq. \ref{eq_hopj32} in main text) are listed 
in Table \ref{tab_hopj32}.

To ensure SU(4) symmetry, we require all fifteen polynomials  $W_{\alpha}=0$ (for $\alpha \neq 0$) in Eq. \ref{eq_loop} of the main text. For generic values
of the hopping parameters, this gives fifteen equations which are solved, resulting in eight independent solutions leading to
eight different SU(4) limits, as discussed in the main text.
The associated plots are given in Fig. \ref{fig:hopping-solutions} of the main text.

\begin{table}
    \centering
    \begin{tabular}{|c|c|c|c|} \hline
     $t^{(\alpha)}_{ij}$    & X-Bond & Y-Bond & Z-Bond \\ \hline\hline 
     $t^{(0)}_{ij}$ & $\frac{1}{3} (2 t_{\pi} + t_{\sigma})$ & $\frac{1}{3} (2 t_{\pi} + t_{\sigma})$ & $\frac{1}{3} (2 t_{\pi} + t_{\sigma})$\\ \hline
     $t^{(1)}_{ij}$ & $-\frac{t_{m}}{\sqrt{3}}$ & $-\frac{t'_{m}}{\sqrt{3}}$ & $-\frac{t'_{m}}{\sqrt{3}}$ \\ \hline
     $t^{(2)}_{ij}$ & $-\frac{t'_{m}}{\sqrt{3}}$ & $-\frac{t_{m}}{\sqrt{3}}$ & $-\frac{t'_{m}}{\sqrt{3}}$ \\ \hline
     $t^{(3)}_{ij}$ & $-\frac{t'_{m}}{\sqrt{3}}$ & $-\frac{t'_{m}}{\sqrt{3}}$ & $-\frac{t_{m}}{\sqrt{3}} $ \\ \hline
     $t^{(4)}_{ij}$ & $\frac{t_{\pi} - t_{\sigma}}{2 \sqrt{3}}$ & $-\frac{t_{\pi} - t_{\sigma}}{2 \sqrt{3}}$ & 0\\ \hline
     $t^{(5)}_{ij}$ & $\frac{t_{\sigma} - t_{\pi}}{6} $ & $\frac{t_{\sigma} - t_{\pi}}{6}$ & $-\frac{(t_{\sigma} - t_{\pi})}{3}$ \\ \hline
    \end{tabular}
    \caption{The amplitudes for different NN hoppings for the $J=3/2$ orbitals as discussed in Eq. \ref{eq_hopj32} of the main text.}
    \label{tab_hopj32}
\end{table}
\section{The Fermionic representation of SU(4) spins}
\label{appen_SU4}

Starting with the $J$-$3/2$ matrices, ($J_x, J_y, J_z$)
\begin{equation}
\begin{tabular}{l}
$J_{x} =
\left(
\begin{array}{cccc}
 0 & 1 & \frac{\sqrt{3}}{2} & 0 \\
 1 & 0 & 0 & \frac{\sqrt{3}}{2} \\
 \frac{\sqrt{3}}{2} & 0 & 0 & 0 \\
 0 & \frac{\sqrt{3}}{2} & 0 & 0 \\
\end{array}
\right)$,
$J_{y} =
\left(
\begin{array}{cccc}
 0 & -i & \frac{i \sqrt{3}}{2} & 0 \\
 i & 0 & 0 & -\frac{i \sqrt{3}}{2} \\
 -\frac{i \sqrt{3}}{2} & 0 & 0 & 0 \\
 0 & \frac{i \sqrt{3}}{2} & 0 & 0 \\
\end{array}
\right),$
$J_{z} =
\left(
\begin{array}{cccc}
 \frac{1}{2} & 0 & 0 & 0 \\
 0 & -\frac{1}{2} & 0 & 0 \\
 0 & 0 & \frac{3}{2} & 0 \\
 0 & 0 & 0 & -\frac{3}{2} \\
\end{array}
\right)$
\end{tabular}
\label{eq_jmat}
\end{equation}

\noindent
One can obtain the five generators of the Euclidean Clifford  Algebra~\cite{PhysRevB.108.245106,PhysRevB.69.235206}, ($\Sigma^1, \cdots, \Sigma^5$) as
\begin{align}
    &\Sigma^{1}=\frac{1}{\sqrt{3}}\{J_{y}, J_{z}\},~~~
     \Sigma^{2}=\frac{1}{\sqrt{3}}\{J_{z}, J_{x}\},~~~
\Sigma^{3}=\frac{1}{\sqrt{3}}\{J_{x}, J_{y}\},~~~
\Sigma^{4}=\frac{1}{\sqrt{3}}(J_{x}^2-J_{y}^2),~~~ \Sigma^{5}=J_{z}^2-\frac{5}{4}\mathbb{I}_4
\label{eq_sigmaj}
\end{align}
where $\mathbb{I}_4$ is the $4\times4$ identity matrix. Notably, $\{\Sigma^{\alpha},\Sigma^{\beta}\}=2\delta_{\alpha\beta}$. 
 From these, one can define another 10 matrices as
\begin{align}
    &\Sigma^{\alpha\beta}=\frac{1}{2i}[\Sigma^{\alpha},\Sigma^{\beta}]
\end{align}

These fifteen $\Sigma$ traceless Hermitian matrices generate the four dimensional representation of SU(4) that satisfy the Lie Algebra~\cite{Haber:2021SU} $[\Sigma^A, \Sigma^B] = i f_{ABC} \Sigma^C$
 where the superscripts $A, B, C$ can take single or double integer indices as described above, and \( f_{ABC} \)-s are the structure constants of SU(4) whose explicit forms can be derived starting from those of the $J$-matrices, $J_i$, given above. Using the above properties, it is fairly easy to show that the operators in Eq. \ref{eq_su4op1} of the main text obey the SU(4) algebra.

\begin{table*}[htb]
\centering
\renewcommand{\arraystretch}{1.3}
\begin{tabular}{|p{0.95\textwidth}|}
\hline
\multicolumn{1}{|c|}{\textbf{Direct-limit}} \\
\hline
\[
g_{1} = g_{2} = g_{3} = g_{4} = \mathbb{I}_{4}
\]
\\
\hline
\multicolumn{1}{|c|}{\textbf{Indirect-limit}} \\
\hline
\begin{minipage}{0.95\textwidth}
\[
\begin{array}{rl}
g_{1} &= -\mathbb{I}_{4},~~~~~~~~~~~~~g_{2} = -\Sigma^1,~~~~~~~g_{3} = \Sigma^2 \Sigma^1 \\
g_{4} &= -\Sigma^3 \Sigma^2 \Sigma^1,~~~~g_{5} = -\Sigma^3 \Sigma^1,~~g_{6} = \Sigma^3 \\
g_{7} &= \Sigma^3 \Sigma^2, ~~~~~~~~~g_{8}= -\Sigma^2
\end{array}
\]
\end{minipage}
\\
\hline
\multicolumn{1}{|c|}{\textbf{$r=-\frac{1}{2}$-limit}} \\
\hline
\begin{minipage}{0.95\textwidth}
\[
\begin{array}{rl}
g_{1} &= e^{-i \frac{2\pi}{3} \Sigma^{45}} e^{-i \frac{3\pi}{2} \Sigma^5},~~~g_{2}= e^{-i \frac{2\pi}{3} \Sigma^{45}},~~~~~~~~~~~~~g_{3}= e^{-i \frac{3\pi}{2} \Sigma^5}, \\
g_{4} &= \mathbb{I}_{4},~~~~~~~~~~~~~~~~~~~~~~~g_{5} = e^{i \frac{2\pi}{3} \Sigma^{45}} e^{-i \frac{3\pi}{2} \Sigma^5},~~~g_{6} = e^{i \frac{2\pi}{3} \Sigma^{45}}, \\
g_{7} &= e^{-i \frac{2\pi}{3} \Sigma^{45}} e^{-i \frac{3\pi}{2} \Sigma^5},~~~g_{8} = e^{-i \frac{2\pi}{3} \Sigma^{45}}, ~~~~~~~~~~~~g_{9} = e^{-i \frac{3\pi}{2} \Sigma^5},\\ \nonumber
g_{10} &= \mathbb{I}_{4}, ~~~~~~~~~~~~~~~~~~~~~~~g_{11}= e^{i \frac{2\pi}{3} \Sigma^{45}} e^{-i \frac{3\pi}{2} \Sigma^5},~~g_{12}= e^{i \frac{2\pi}{3} \Sigma^{45}}.
\end{array}
\]
\end{minipage}
\\
\hline
\end{tabular}
\caption{Representations of $g_i$ matrices in different limits. The matrices $\Sigma$ are $4\times 4$ SU(4) generators. The site-indices ($i$) for $g_{i}$ correspond to sites presented in Fig~\ref{fig:sites_presentations} in the main text.}
\label{gi_table}
\end{table*}
\section{Rotation of SU(4) operators from global to local basis}\label{app:global_to_local_rot}

The $J=3/2$ fermions, $\psi_{i}$, in global basis are related by a unitary rotation $g_{i}$ in the local fermions~$\phi_{i}$ [cf. Eq.~\ref{eq_gphi} in main text]. The list of $g_{i}$ for the three limits is listed in Table~\ref{gi_table}. This relation connects the local SU(4) operator $O^{(a)}_{i}$ (cf. Eq. \ref{eq_su4op1} in main text) with the global version of SU(4) operator given by, 
\begin{align}
    \tilde{O}^{(a)}_{i}=\psi^{\dagger}_{i}\Sigma^{a} \psi_{i}
    \label{eq_globalop}
\end{align}


In the local basis, this SU(4) operator can be expressed as (using Eq.~\ref{eq_gphi} in main text):
\begin{align}
    \tilde{O}^{(a)}_{i}=&\phi^{\dagger}_{i}g_{i}^{\dagger}\Sigma^{a} g_{i}\phi_{i}=\sum_{{b}} u^{(i)}_{ab} O^{(b)}_{i}=\sum_{b} u^{(i)}_{ab} (\phi^{\dagger}_{i} \Sigma^{b} \phi_{i})=\phi^{\dagger}_{i}(\sum_{b} u^{(i)}_{ab} \Sigma^{b}) \phi_{i}  \nonumber\\
\end{align}

Thus, this implies that the relation between the $4\times 4$ matrices in the global and local bases is given by
\begin{align}
    g^{\dagger}_{i}\Sigma^{a}g_{i}=\sum_{b} u^{(i)}_{ab} \Sigma^{b} 
\end{align}

where 
\begin{align}
   u^{(i)}_{ab}=&\frac{Tr(g_i^{\dagger}.\Sigma^{a}.g_i.\Sigma^{b})}{Tr(\Sigma^{b}.\Sigma^{b})}=\frac{1}{4}Tr(g_i^{\dagger}.\Sigma^{a}.g_i.\Sigma^{b})
\end{align}

Thus, this connect the $O^{(a)}_{i}$ with $\tilde{O}^{(a)}_{i}$ as 
\begin{align}
    \tilde{O}^{(a)}_{i}=\sum_{b=1}^{15} u^{(i)}_{ab} O^{(b)}_{i}
\label{eq_globallocal}
\end{align}

here, {$u^{(i)}_{ab}$ is $15 \times15$ matrix acting on of 15-dimensional SU(4) operators basis $(O^{(1)}_{i}, O^{(2)}_{i}, O^{(3)}_{i}, .. O^{(15)}_{i})$. The list of $u^{(i)}$ for all three limits has been summarised in Table~\ref{ui_table}.

\begin{table*}[htb]
\centering
\renewcommand{\arraystretch}{1.3}
\begin{tabular}{|p{0.95\textwidth}|}
\hline
\multicolumn{1}{|c|}{\textbf{Direct-limit}} \\
\hline
\[
u^{(1)} = u^{(2)} = u^{(3)} = u^{(4)} = \mathbb{I}_{15}
\]
\\
\hline
\multicolumn{1}{|c|}{\textbf{Indirect-limit}} \\
\hline
\begin{minipage}{0.95\textwidth}
\[
\begin{array}{rl}
u^{(1)} = & \mathbb{I}_{15} \\
u^{(2)} = & \mathbb{I}_{1}\oplus (-\mathbb{I}_{8}) \oplus \mathbb{I}_{6} \\
u^{(3)} = & (-\mathbb{I}_{2}) \oplus \mathbb{I}_{4} \oplus (-\mathbb{I}_{6}) \oplus \mathbb{I}_{3} \\
u^{(4)} = & \mathbb{I}_{3} \oplus (-\mathbb{I}_{2}) \oplus \mathbb{I}_{2} \oplus (-\mathbb{I}_{2}) \oplus \mathbb{I}_{1} \oplus (-\mathbb{I}_{4}) \oplus \mathbb{I}_{1} \\
u^{(5)} = & (-\sigma_z)\oplus(-\mathbb{I}_{1}) \oplus \mathbb{I}_{2} \oplus (-\sigma_z)\oplus (-\mathbb{I}_{3}) \oplus \mathbb{I}_{2} \oplus (-\mathbb{I}_{2}) \oplus \mathbb{I}_{1} \\
u^{(6)} = & (-\mathbb{I}_{2}) \oplus \sigma_z \oplus (-\sigma_z) \oplus (-\sigma_z) \oplus \sigma_z \oplus \mathbb{I}_{2} \oplus (-\mathbb{I}_{2}) \oplus \mathbb{I}_{1} \\
u^{(7)} = & \mathbb{I}_{1} \oplus (-\mathbb{I}_2) \oplus \mathbb{I}_2 \oplus (-\mathbb{I}_2) \oplus \mathbb{I}_3 \oplus (-\mathbb{I}_4) \oplus \mathbb{I}_{1} \\
u^{(8)} = & (-\sigma_z) \oplus (-\mathbb{I}_4) \oplus \mathbb{I}_3 \oplus (-\mathbb{I}_3) \oplus \mathbb{I}_3
\end{array}
\]
\end{minipage}
\\
\hline
\multicolumn{1}{|c|}{\textbf{$r=-\frac{1}{2}$-limit}} \\
\hline
\begin{minipage}{0.95\textwidth}
\[
\begin{array}{rl}
&u^{(1)}=u^{(7)}=  
(-\mathbb{I}_3) \oplus
\begin{pmatrix}
\tfrac{1}{2} & \tfrac{\sqrt{3}}{2} \\
\tfrac{\sqrt{3}}{2} & -\tfrac{1}{2}
\end{pmatrix}
\oplus \mathbb{I}_2 \oplus
\begin{pmatrix}
-\tfrac{1}{2} & -\tfrac{\sqrt{3}}{2} \\
-\tfrac{\sqrt{3}}{2} & \tfrac{1}{2}
\end{pmatrix}
\oplus \mathbb{I}_1 \oplus
\begin{pmatrix}
-\tfrac{1}{2} & -\tfrac{\sqrt{3}}{2} \\
-\tfrac{\sqrt{3}}{2} & \tfrac{1}{2}
\end{pmatrix} \oplus
\begin{pmatrix}
-\tfrac{1}{2} & -\tfrac{\sqrt{3}}{2} \\
-\tfrac{\sqrt{3}}{2} & \tfrac{1}{2}
\end{pmatrix}
\oplus (-\mathbb{I}_1) \\
&u^{(2)}=u^{(8)}= 
\mathbb{I}_3 \oplus
\begin{pmatrix}
-\tfrac{1}{2} & \tfrac{\sqrt{3}}{2} \\
-\tfrac{\sqrt{3}}{2} & -\tfrac{1}{2}
\end{pmatrix}
\oplus \mathbb{I}_2 \oplus
\begin{pmatrix}
-\tfrac{1}{2} & \tfrac{\sqrt{3}}{2} \\
-\tfrac{\sqrt{3}}{2} & -\tfrac{1}{2}
\end{pmatrix}
\oplus \mathbb{I}_1 \oplus
\begin{pmatrix}
-\tfrac{1}{2} & \tfrac{\sqrt{3}}{2} \\
-\tfrac{\sqrt{3}}{2} & -\tfrac{1}{2}
\end{pmatrix}
\oplus
\begin{pmatrix}
-\tfrac{1}{2} & \tfrac{\sqrt{3}}{2} \\
-\tfrac{\sqrt{3}}{2} & -\tfrac{1}{2}
\end{pmatrix}
\oplus \mathbb{I}_1 \\
&u^{(3)}=u^{(9)}=(-\mathbb{I}_4) \oplus (\mathbb{I}_{4}) \oplus (-\sigma_{z}) \oplus (\sigma_z) \oplus \sigma_{z} \oplus (-\mathbb{I}_{1}) \\
&u^{(4)} = u^{(10)} = \mathbb{I}_{15} \\
&u^{(5)} = u^{(11)} = 
(-\mathbb{I}_3) \oplus
\begin{pmatrix}
\tfrac{1}{2} & -\tfrac{\sqrt{3}}{2} \\
-\tfrac{\sqrt{3}}{2} & -\tfrac{1}{2}
\end{pmatrix}
\oplus \mathbb{I}_2 \oplus
\begin{pmatrix}
-\tfrac{1}{2} & \tfrac{\sqrt{3}}{2} \\
\tfrac{\sqrt{3}}{2} & \tfrac{1}{2}
\end{pmatrix}
\oplus \mathbb{I}_1 \oplus
\begin{pmatrix}
-\tfrac{1}{2} & \tfrac{\sqrt{3}}{2} \\
\tfrac{\sqrt{3}}{2} & \tfrac{1}{2}
\end{pmatrix}
 \oplus
\begin{pmatrix}
-\tfrac{1}{2} & \tfrac{\sqrt{3}}{2} \\
\tfrac{\sqrt{3}}{2} & \tfrac{1}{2}
\end{pmatrix}
\oplus (-\mathbb{I}_1) \\
&u^{(6)}=u^{(12)}= 
\mathbb{I}_3 \oplus
\begin{pmatrix}
-\tfrac{1}{2} & -\tfrac{\sqrt{3}}{2} \\
\tfrac{\sqrt{3}}{2} & -\tfrac{1}{2}
\end{pmatrix}
\oplus \mathbb{I}_2 \oplus
\begin{pmatrix}
-\tfrac{1}{2} & -\tfrac{\sqrt{3}}{2} \\
\tfrac{\sqrt{3}}{2} & -\tfrac{1}{2}
\end{pmatrix}
\oplus \mathbb{I}_1 
\oplus
\begin{pmatrix}
-\tfrac{1}{2} & -\tfrac{\sqrt{3}}{2} \\
\tfrac{\sqrt{3}}{2} & -\tfrac{1}{2}
\end{pmatrix}
\oplus
\begin{pmatrix}
-\tfrac{1}{2} & -\tfrac{\sqrt{3}}{2} \\
\tfrac{\sqrt{3}}{2} & -\tfrac{1}{2}
\end{pmatrix}
\oplus \mathbb{I}_1
\end{array}
\]
\end{minipage}
\\
\hline
\end{tabular}
\caption{Block-diagonal representations of the matrices $u^{(i)}$ in various limits, shown as direct sums where $\mathbb{I}_n$ is the $n\times n$ identity matrix and  $\sigma_z$ is the $z$-Pauli matrix. The site-indices ($i$) for $u^{(i)}$ correspond to sites presented in \Fig{fig:sites_presentations} in the main text.}
\label{ui_table}
\end{table*}

\section{The SU(4) spin Hamiltonian in global basis}
\label{appen_spin}

To understand the difference in the implementation of the microscopic symmetries on the three SU(4) limits, it is instructive to derive the spin Hamiltonian in the global basis starting from Eq. \ref{eq_j32hubbard}. To order $\sim t^2/U$, we get :


\begin{align}
H_{\text{eff}} &= \dfrac{6}{U} \sum_{\langle ij \rangle}\Bigg( 
\sum_{a=1}^{15} \dfrac{1}{4} (t^{(0)}_{ij})^2 \tilde{O}_{i}^{(a)} \tilde{O}_{j}^{(a)} +  \sum_{\alpha=1}^{5} \sum_{a,b=1}^{15} \dfrac{t^{(0)}_{ij} t^{(\alpha)}_{ij}}{16} \, \text{Tr}\left( \Sigma^{\alpha} \{\Sigma^{a}, \Sigma^{b}\} \right) \tilde{O}_{i}^{(a)} \tilde{O}_{j}^{(b)}  \nonumber \\
&~~~~~~~~~~~~~~~~+ \sum_{\alpha,\alpha'=1}^{5} \sum_{a,b=1}^{15}  \dfrac{t^{(\alpha)}_{ij} t^{(\alpha')}_{ij}}{16} \, \text{Tr}\left( \Sigma^{\alpha} \Sigma^{b} \Sigma^{\alpha'} \Sigma^{a} \right) \tilde{O}_{i}^{(a)} \tilde{O}_{j}^{(b)}
\Bigg),
\label{eq:Heff}
\end{align}

where the symbols $t^{(0)}_{ij}$, $t^{(\alpha)}_{ij}$ etc. are discussed in  \App{appen_electronic}. The first term in $H_{\text{eff}}$ is SU(4) symmetric in the global basis and corresponds to the SU(4) Heisenberg antiferromagnet described by case-I where $t^{(\alpha)}_{ij}\propto \delta_{\alpha,0}$ such that the antiferromagnetic coupling constant $\mathcal{J}= 3t_{\sigma}^2 / 2U$. For case-II, {\it i.e.}, the indirect limit, only $t_m\neq 0$ such that we use  \Eq{eq_globallocal} and Table~\ref{ui_table} to obtain the SU(4) Hamiltonian in the local basis (Eq. \ref{eq_su4afm} in main text) with $\mathcal{J}=t^{2}_{m}/2U$. Finally,  for case-III, we have  $t_{\pi}/t_{\sigma} = -1/2,\ t_{m} = t'_{m} = 0)$ such that a more involved set of local rotations is needed. This is given by \Eq{eq_globallocal} and Table~\ref{ui_table} whence we get to \Eq{eq_su4afm} in the main text with $\mathcal{J} = 3t^{2}_{\sigma}/8U$. 

%
\section{The SU(4) Heisenberg model and mean field decoupling}
\label{appen_Heisen}

Using the parton decomposition, we rewrite the emergent Heisenberg Hamiltonian in \Eq{eq_su4afm} of the main text as
\begin{align}
    H=-\mathcal{J}\sum_{\langle ij\rangle} :\chi_{ij}^\dagger\chi_{ij} :,
    \label{eq_sunform}
\end{align}
where $:O:$ denotes normal ordering and \begin{align}
    \chi_{ij}=f_{i\alpha}^\dagger f_{j\alpha}
\end{align}
is the SU(N) bond singlet and the dynamics is constrained by \Eq{eq_su4const} of the maintext. For the magnetic unit-cell containing $M=4,8,12$ sites (see \Fig{fig:sites_presentations}) we define the Fourier modes for each fermion-flavor $\alpha=\{1,..,\mathcal{N}_\mathit{f}\}$, as

\begin{align}
     f_{{\bf R}_{i}+{\bm{\delta}}_{a}, \alpha}=&\frac{1}{\sqrt{N}}~\sum_{\bf k\in MBZ}~e^{-i{\bf k}.({\bf R}_{i}+{\bm{\delta}}_a)}f_{{\bf k}, a,\alpha},  
     \label{eq_ft}
\end{align}     
where ${\bf R_i}$ is the Bravais lattice vector for the $i$-th magnetic unit-cell, $a=\{1,\cdots, M\}$ denotes the sub-lattice within the unit-cell,  $\bm{\delta}_a$ is the internal coordinate of the $a$-th sub-lattice atom, and the ${\bf k}$-sum iterates over the magnetic Brillouin Zone (MBZ), which, given the intricacies, we briefly discuss now.

\paragraph*{2-site Unit Cell :} The honeycomb lattice has a two site unit cell with atoms located at {$\boldsymbol{\delta}_{1}=(0,0)$, and $\boldsymbol{\delta}_{2}=(\frac{1}{2\sqrt{3}}, \frac{1}{2} )$} with lattice vectors ${\bf r}_{\bf n}=n_1 {\bf a}+n_2{\bf b}$ where ${\bf n}=(n_1, n_2)$ with $n_1, n_2\in \mathbb{Z}$ and 
\begin{align}
\mathbf{a}= \left( \frac{\sqrt{3}}{2}, -\frac{1}{2} \right)~~~~~~~~~~~~~~~~\mathbf{b}= \left( 0, 1 \right)
\end{align}
The corresponding reciprocal lattice vectors are ${\bf G}_{\bf n}=n_1{\bf a}^*+n_2{\bf b}^*$ with 
\begin{align}
\mathbf{a}^*= \left( \frac{2\pi}{\sqrt{3}}, -\frac{2\pi}{\sqrt{3}} \right),~~~~~~~~~~~~~\mathbf{b}^* = \left( 0, \frac{2\pi}{3} \right)
\end{align}
The first BZ is shown in Fig. \ref{fig:parton_mean_field}(d, black).

\paragraph*{4-site Unit Cell :} The unit cell is shown in Fig. \ref{fig:sites_presentations}(a) of the main text, and the position of the sublattices is
\begin{align}
    \boldsymbol{\delta}_1=(0, 0),~~\boldsymbol{\delta}_{2}=(\frac{1}{2\sqrt{3}},~~\frac{1}{2}), \boldsymbol{\delta}_{3}=(\frac{\sqrt{3}}{2},~~\frac{1}{2}), \boldsymbol{\delta}_{4}&=(\frac{2}{\sqrt{3}}, 0)
\end{align}

The lattice vectors are given by ${\bf R}_i=i_1{\bf a}+i_2{\bf b}$ where $i=(i_1,i_2)$ with $i_1,i_2\in\mathbb{Z}$ and 
\begin{align}
\mathbf{a}= \left( \sqrt{3}, 0 \right)~~~~~~~~\mathbf{b} &= \left( 0, 1 \right).
\end{align}
The corresponding reciprocal lattice vectors are given by ${\bf G}_i=i_1{\bf a}^*+i_2{\bf b}^*$ with 
\begin{align}
\mathbf{a}^*= \left( \frac{2\pi}{\sqrt{3}}, 0 \right),~~~~~~~~~~~~~~~~
\mathbf{b}^*= \left( 0, 2\pi \right)
\end{align}
The first BZ is shown in Fig. \ref{fig:parton_mean_field}(d, blue).

\paragraph*{8-site Unit Cell} The unit cell is shown in Fig. \ref{fig:sites_presentations}(b) of the main text, and the position of the sublattices is
\begin{align}
    \boldsymbol{\delta}_1=(0, 0), ~~\boldsymbol{\delta}_{2}=(\frac{1}{2\sqrt{3}}, \frac{1}{2}), ~~\boldsymbol{\delta}_{3}=(\frac{\sqrt{3}}{2}, \frac{1}{2}),   ~~\boldsymbol{\delta}_{4}=(\frac{2}{\sqrt{3}}, 0) \nonumber \\
    \boldsymbol{\delta}_{5}=(0, 1), ~~\boldsymbol{\delta}_{6}=(\frac{1}{2\sqrt{3}}, \frac{3}{2}),~~\boldsymbol{\delta}_{7}=(\frac{\sqrt{3}}{2}, \frac{3}{2}),   ~~\boldsymbol{\delta}_{8}=(\frac{2}{\sqrt{3}}, 1)
\end{align}
The lattice vectors are given by ${\bf R}_i=i_1{\bf a}+i_2{\bf b}$ where $i=(i_1,i_2)$ with
\begin{align}
\mathbf{a} &= \left( \sqrt{3}, -1 \right),~~~\mathbf{b}= \left( 0, 2 \right)
\end{align}
The corresponding reciprocal lattice vectors are given by ${\bf G}_i=i_1{\bf a}^*+i_2{\bf b}^*$ with 
\begin{align}
\mathbf{a}^*= \left( \frac{2\pi}{\sqrt{3}}, 0 \right),~~~~~~~~ \mathbf{b}^* = \left( \frac{\pi}{\sqrt{3}}, \pi \right)
\end{align}
The first BZ is shown in Fig. \ref{fig:parton_mean_field}(d, red).

\paragraph*{12-site Unit Cell}: The unit cell is shown in Fig. \ref{fig:sites_presentations}(c) of the main text, and the position of the sublattices is

\begin{align}
    \boldsymbol{\delta}_1=&(0, 0),~~\boldsymbol{\delta}_{2}=(\frac{1}{2\sqrt{3}}, \frac{1}{2}),~~\boldsymbol{\delta}_{3}=(\frac{\sqrt{3}}{2}, \frac{1}{2}),~~\boldsymbol{\delta}_{4}=(\frac{2}{\sqrt{3}}, 0) \nonumber \\
    \boldsymbol{\delta}_{5}=& \left( \frac{3\sqrt{3}}{2}, \frac{5}{2} \right),~~\boldsymbol{\delta}_{6} = \left( \frac{7}{2\sqrt{3}}, \frac{5}{2} \right),   ~~\boldsymbol{\delta}_{7} =\left(\frac{\sqrt{3}}{2}, \frac{3}{2}\right), ~~\boldsymbol{\delta}_{8} =\left(\frac{2}{\sqrt{3}}, 2\right) \nonumber \\
    ~~\boldsymbol{\delta}_{9} =&\left(\sqrt{3}, 2\right), ~~\boldsymbol{\delta}_{10}=\left(\frac{7}{2\sqrt{3}}, \frac{3}{2}\right),~~\boldsymbol{\delta}_{11} =\left(\sqrt{3}, 1\right), ~~\boldsymbol{\delta}_{12} =\left(\frac{2}{\sqrt{3}}, 1\right)
\end{align}
The lattice vectors are given by ${\bf R}_i=i_1{\bf a}+i_2{\bf b}$ where $i=(i_1,i_2)$ with
\begin{align}
\mathbf{a} &= \left( \sqrt{3}, 0 \right) \nonumber \\
\mathbf{b} &= \left( \sqrt{3}, 3 \right)
\end{align}
The corresponding reciprocal lattice vectors are given by ${\bf G}_i=i_1{\bf a}^*+i_2{\bf b}^*$ with 
\begin{align}
\mathbf{a}^*= \left( \frac{2\pi}{\sqrt{3}}, -\frac{2\pi}{\sqrt{3}} \right),~~~~~~~~
\mathbf{b}^* = \left( 0, \frac{2\pi}{3} \right)
\end{align}
The first BZ is shown in Fig. \ref{fig:parton_mean_field}(d, green).
\subsection*{The parton band-structure}

The Mean field Hamiltonian is then given by 
\begin{align}
    H_{MF}=\frac{\mathcal{J}N_b}{N_f}\chi^2-\frac{\chi\mathcal{J}}{N_{f}}\sum_{\alpha}\sum_{\bf k\in MBZ}{\bf f}^\dagger_{\bf k, \alpha}~\mathcal{H}({\bf k})~{\bf f}_{\bf k, \alpha}
    \label{eq:app_HMF}
\end{align}
where ${\bf f}_{\bf k}\equiv\left[f_1({\bf k})\cdots f_M({\bf k})\right]^T$ is an $M$-component fermion annihilation operator (Eq. \ref{eq_ft}), and $\mathcal{H}({\bf k})$ is a ${\bf k}$-dependent $M\times M$ matrix that is same for all the flavours due to the SU(4) symmetry. The resultant parton band structure obtained from diagonalising $\mathcal{H}({\bf k})$ is shown in \Fig{fig:parton_mean_field}. The mean-field parameter $\chi$ is self-consistently determined from the four-site calculation as 
\begin{align}
    \chi=\frac{1}{2}\int_{\bf MBZ} \frac{d^2{\bf k}}{(2\pi)^2}~\varepsilon_4({\bf k})
\end{align}
where $\varepsilon_4({\bf k})$ is the dispersion of the occupied band for the four-site calculation.
\begin{figure}[htb]
    \centering
    \includegraphics[width=0.8\linewidth]{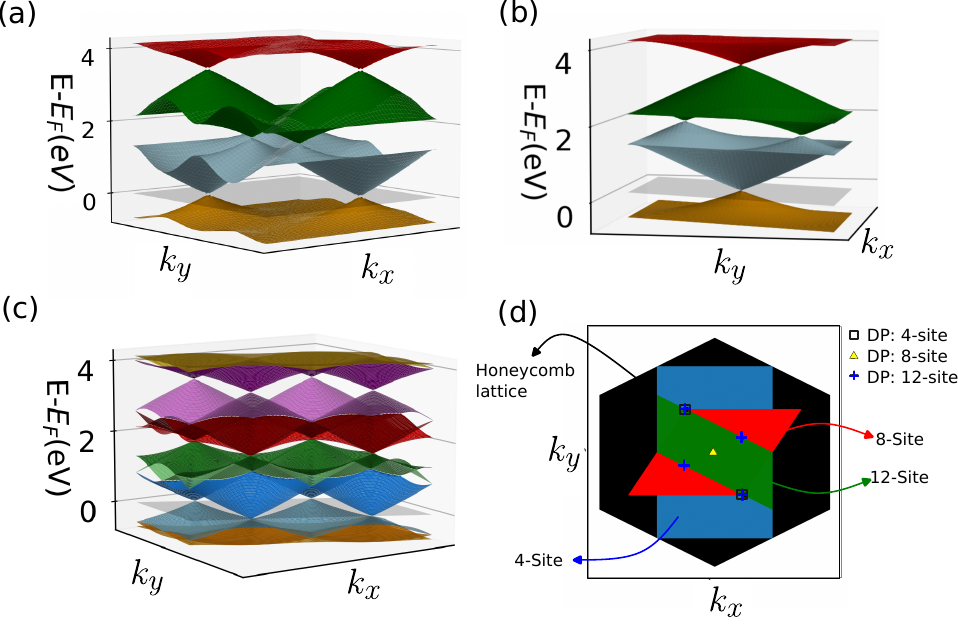}
    \caption{Parton mean-field band structure. Panels (a), (b), and (c) show the band structures of the parton mean-field Hamiltonian for the 4-site, 8-site, and 12-site magnetic unit cells, respectively, each plotted within its corresponding magnetic Brillouin zone (MBZ). The Fermi energy is indicated with a shaded plane. Panel (d) compares these MBZs with the BZ of the original honeycomb lattice. The positions of the Dirac points (DPs) are also highlighted: for the 4-site case, they appear at $k = \pm(\pi/2\sqrt{3}, -\pi/2)$; for the 8-site case, at $\Gamma$; and for the 12-site case, at $k=\{\left( -\pi/2\sqrt{3}, \pi/{2} \right)$, $\left( \pi/2\sqrt{3}, \pi/6 \right)$\}. }
    \label{fig:parton_mean_field}
\end{figure}


\section{Dynamical structure factor calculations}
\label{appen_sf}
The dynamical $J=3/2$ dipole structure factor given by Eq.~\ref{eq:dyn_struct_fac} of the main text can be calculated starting with the rewriting of the $J=3/2$ dipole matrices (Eq. \ref{eq_jmat}) in terms of the SU(4) generators (and inverting Eq. \ref{eq_sigmaj}), as
\begin{align}
J_{x} =\frac{\sqrt{3}}{2}\Sigma^{15}-\frac{1}{2}(\Sigma^{23}-\Sigma^{14}), ~~~J_{y}=&-\frac{\sqrt{3}}{2} \Sigma^{25}+\frac{1}{2}(\Sigma^{13}+\Sigma^{24}), ~~~J_{z}=-\Sigma^{34}-\frac{1}{2}\Sigma^{12}.
\label{eq:J3by2}
\end{align}
The forms of the $\Sigma$ matrices, in turn, are defined in \ref{appen_SU4} with the corresponding global SU(4) operators, $\tilde{O}^{(p)}$, being obtained using \Eq{eq_globalop}, such that 
    \begin{align}
    J^\sigma_{i} &= \sum_{p=1}^{15} c_{\sigma p} \tilde{O}^{(p)}_{i} ~~~\sigma=x,y,z
    \end{align}
with the constants $c_{\sigma p}$ obtained from Eq. \ref{eq:J3by2}. The $\tilde{O}^{(p)}$ (global basis) can be expressed in terms of the local operators $O^{(p)}$ using Eq. \ref{eq_globallocal} to express the dipole correlations as
\begin{align}
\langle {\bf J}_i(t)\cdot {\bf J}_j(0)\rangle=\sum_{p,q,r,s=1}^{15} \sum_{\sigma}& c_{\sigma p} c_{\sigma q}[u^{(i)}]_{pr}[u^{(j)}]_{qs} 
~ \langle O^{(r)}_{i}(t) O^{(s)}_{j}(0) \rangle, 
\end{align}
where the symbols $u^{(i)}$ etc. are defined in \ref{app:global_to_local_rot}

To proceed further, we find it convenient to use the imaginary time formalism $t\rightarrow -i\tau$ where $\tau\in[0,1/T]$ is the imaginary or Euclidean time (and $T$ is the temperature) to obtain 
\begin{align}
     \langle {\bf J}_i(\tau)\cdot {\bf J}_j(0)\rangle=\sum_{\alpha \beta \gamma \delta} ^{4}~\Xi_{\alpha \beta \gamma \delta}^{ij}~ \langle~f^{\dagger}_{i\alpha}(\tau)  f_{i \beta}(\tau)~f^{\dagger}_{j\gamma}(0) f_{j \delta}(0)~\rangle
     \label{Eq:Xi}
\end{align}
where the SU(4) generators are expressed in terms of the fermions using Eq. \ref{eq_su4op1} of the main text and 
\begin{align}
    \Xi_{\alpha \beta \gamma \delta}^{ij}= \sum_{pqrs=1}^{15} \sum_{\sigma} c_{\sigma p}[u^{(i)}]_{pr} c_{\sigma q}[u^{(j)}]_{qs}\Sigma^{r}_{\alpha \beta}~\Sigma^{s}_{\gamma \delta},
\end{align}
is the form factor with $\alpha, \beta, \gamma$ and $\delta$ being the fermion flavor indices.

At this point, we resolve the site indices $i$, $j$ in terms of the Bravais lattice vectors ($\mathbf{R_i}$, $\mathbf{R_j}$), spanning the magnetic unit cell (defined in Section \ref{appen_Heisen}), and their relative sublattice coordinates ($\bm{\delta}_{\M}$, $\bm{\delta}_{\N}$), respectively and rewrite 
\begin{align}
 &\langle {\bf J}_{{\bf R}_i+\bm{\delta}_\M}(\tau)\cdot {\bf J}_{{\bf R}_j+\bm{\delta}_\N}(0)\rangle=\sum_{\alpha \beta \gamma \delta} ^{4}~\Xi^{\M,\N}_{\alpha \beta \gamma \delta}\langle~f^{\dagger}_{\mathbf{R}_{i}+\bm{\delta}_{\M},\alpha}(\tau)  f_{\mathbf{R}_{i}+\bm{\delta}_{\M}, \beta}(\tau)~f^{\dagger}_{\mathbf{R}_{j}+\bm{\delta}_{\N},\gamma}(0) f_{\mathbf{R}_{j}+\bm{\delta}_{\N} \delta}(0)~\rangle
\end{align} 
where we have used the fact that due to translation symmetry associated with the magnetic unit cell
\begin{align}
    &~~~~~~~~~~~~\Xi^{\R_{i}+\bdelta_{\M}, \R_{j}+\bdelta_{\N}}_{\alpha \beta \gamma \delta}=\Xi^{\bdelta_{\M}, \bdelta_{\N}}_{\alpha \beta \gamma \delta}\equiv\Xi^{\M, \N}_{\alpha \beta \gamma \delta}    
\end{align} 

Expressing the Bravais lattice and internal-coordinate-resolved fermionic operators in terms of the Fourier modes (defined in \Eq{eq_ft}), and simplifying the resulting delta functions, we get
\begin{align}
    \langle {\bf J}_{{\R}_i+\bdelta_\M}(\tau)\cdot {\bf J}_{\R_j+\bdelta_\N}(0)\rangle=\frac{1}{N}\sum_{\mathbf{q}} I_{\M,\N}(\mathbf{q},\tau) e^{i \mathbf{q} \cdot (\R_i+\bdelta_\M - \R_j-\bdelta_\N)},
\end{align}
where 

\begin{align}
 &I_{\M,\N}(\mathbf{q},\tau)=\frac{1}{N}~\sum_{k_1, k_2} ~~ \sum_{\alpha \beta \gamma \delta}^{4} \Xi^{\M, \N}_{\alpha \beta \gamma \delta}~\langle f^{\dagger}_{\mathbf{k}_1+\mathbf{q}, \M \alpha}(\tau)  f_{{\mathbf{k}_1}, \M \beta}(\tau) f^{\dagger}_{\mathbf{k}_2, \N\gamma}(0) f_{{\mathbf{k}_2}+\mathbf{q}, \N\delta}(0) \rangle. \label{eq:app_Iq}
\end{align}
The quantity $I_{\M,\N}(\bm{q},\tau)$ will ultimately be used determine the dynamical dipole structure factor $S({\bm q}, \omega)$.
The expectation of the 4-fermion term in \Eq{eq:app_Iq} can be calculated at the mean-field level, by first expanding the $f$ operators in terms of band eigenmodes, $\Phi$, obtained by diagonalising the mean-field Hamiltonian in \Eq{eq:app_HMF}
\begin{align}
f_{\alpha N}(\mathbf{k})=&\sum_{a=1}\Gamma(\mathbf{k})_{Na}\Phi_{\alpha a}(\mathbf{k})
\end{align}
and then inserting the resulting expression into \Eq{eq:app_Iq} to obtain
 \begin{align}
   &I_{\M,\N}(\mathbf{q}, \tau)=\frac{1}{N}~\sum_{k_1, k_2} ~~ \underbrace{ \sum_{\alpha \beta \gamma \delta}^{4} \Xi^{\M, \N}_{\alpha \beta \gamma \delta}}_{\kappa_{\M\N}}~~
    \sum_{a, b, c, d}\Gamma(\mathbf{k_1}+\mathbf{q})^{*}_{\M a} \Gamma(\mathbf{k_1})_{\M b} \Gamma(\mathbf{k_2})^{*}_{\N c} \Gamma(\mathbf{k_2+q})_{\N d} \nonumber\\
    &~~~~~~~~~~~~~~~~~~~~~~~~~~~~~~~~~~~~~~~~~~~~~~~~~~~~~~~~~~~~~~~~~~~~~~~~~~\times~\langle ~\Phi^{\dagger}_{\alpha a}(\mathbf{k_1+q}, \tau) ~\Phi_{\beta b}(\mathbf{k_1}, \tau)~\Phi^{\dagger}_{\gamma c}(\mathbf{k_2})~\Phi_{\delta d}(\mathbf{k_2+q}) \rangle,
    \label{eq:app_Iq2}
\end{align}
where we defined the symbol $\kappa_{\M\N}$ for notational simplicity.
Next, the 4-fermion correlator involving $\Phi$s is decoupled using Wick's theorem and replacing the resulting two-point fermion correlators using the parton mean field theory, {\it i.e.},
\begin{align}
   &\langle ~\Phi^{\dagger}_{\alpha a}(\mathbf{k_1+q}, \tau) \Phi_{\beta b}(\mathbf{k_1}, \tau)~\Phi^{\dagger}_{\gamma c}(\mathbf{k_2})~\Phi_{\delta d}(\mathbf{k_2+q}) \rangle \nonumber\\
    =&~e^{\left(\E_{a}(\mathbf{k_1+q})-\E_{b}(\mathbf{k_1})\right)\tau} 
   \langle ~\Phi^{\dagger}_{\alpha a}(\mathbf{k_1+q}) ~\Phi_{\beta b}(\mathbf{k_1})~\Phi^{\dagger}_{\gamma c}(\mathbf{k_2})~\Phi_{\delta d}(\mathbf{k_2+q}) \rangle\\
    =&~e^{\left(\E_{a}(\mathbf{k_1+q}) - \E_{b}(\mathbf{k_1})\right)\tau}\nonumber \\
    &\times\Big(\underbrace{\langle \Phi_{\alpha a}^{\dagger}(\mathbf{k_1+q}) \Phi_{\beta b}(\mathbf{k_1}) \rangle}_{ \delta_{\alpha\beta} \delta_{ab} \delta(\mathbf{k_1}, \mathbf{k_1+q}) n_{F}(\E_{a}(\mathbf{k_1+q}))} \underbrace{\langle \Phi^{\dagger}_{\gamma c}(\mathbf{k_2}) \Phi_{\delta d}(\mathbf{k_2+q}) \rangle}_{ \delta_{\gamma \delta} \delta_{cd} \delta(\mathbf{k_2}, \mathbf{k_2+q}) n_{F}(\E_{c}(\mathbf{k_2}))}
    + 
\underbrace{\langle \Phi_{\alpha a}^{\dagger}(\mathbf{k_1+q}) \Phi_{\delta d}(\mathbf{k_2+q}) \rangle}_{\delta_{\alpha\delta}\delta_{ad}\delta(\mathbf{k_{1},k_{2}})n_{F}(\E_{a}(\mathbf{k_{1}+q}))} \underbrace{\langle \Phi_{\beta b}(\mathbf{k_1}) \Phi_{\gamma c}^{\dagger}(\mathbf{k_2}) \rangle}_{\delta_{\beta\gamma}\delta_{bc}\delta(\mathbf{k_{1},k_{2}})\left(1-n_{F}(\E_{b}(\mathbf{k_{1}}))\right)} 
\Big), \nonumber
\end{align}
{where we have introduced the fermi-function $n_F(x)=1/(\exp(x)+1)$ and $\E_{a,b}(\k)$ are the dispersion of the eigenmodes.} 
{Inserting the Wick contracted expression for the 4-fermion correlator into to expression for $I_{\M,\N}(\bf{q},\tau)$} in \Eq{eq:app_Iq2} we get, for $\mathbf{q} \neq 0$, 
\begin{align}
 I_{\M,\N}(\mathbf{q}, \tau)
 &=\frac{1}{N}~\sum_{\k} ~{\kappa_{\M\N}}\times \sum_{a, b=1}\Gamma(\mathbf{k}+\mathbf{q})^{*}_{\M a} \Gamma(\mathbf{k})_{\M b} \Gamma(\mathbf{k})^{*}_{\N b} \Gamma(\mathbf{k+q})_{\N a} \nonumber\\
 &~~~~~~~~~~~~~~~~~~~~~~~~~~~~~~~~~~~~~~~~~~~
 \times \left[ e^{\left(\E_{a}(\mathbf{k+q})-\E_{b}(\mathbf{k})\right)\tau} n_{F}(\E_a(\mathbf{k+q}))(1-n_{F}(\E_b(\mathbf{k_1}))\right]. \nonumber
\end{align}
Moving to Matsubara frequencies  $\Omega_{n} = 2\pi n T$ ($n\in \mathbb{Z}$) by taking the Fourier transform of the  $\tau$ term to find
\begin{align}
&I_{\M,\N}(\mathbf{q}, i\Omega_{n})= \kappa_{\M\N} \cdot \frac{1}{N} \sum_{\mathbf{k} \in \text{MBZ}} \sum_{a,b=1} 
\Gamma(\mathbf{k} + \mathbf{q})_{\M a}^{*} \Gamma(\mathbf{k})_{\M b} \Gamma(\mathbf{k})_{\N b}^{*} \Gamma(\k + \mathbf{q})_{\N a} \left[ \frac{n_F(\E_a(\mathbf{k} + \mathbf{q})) \left(1 - n_F(\E_b(\mathbf{k}))\right)}{i\Omega_n + \left(\E_a(\mathbf{k} + \mathbf{q}) - \E_b(\mathbf{k})\right)} \right].
\end{align}
Finally, going from Matsubara frequency to real frequency $\omega$ via analytical continuation, we have
\begin{align}
    I_{\M,\N}(\mathbf{q}, \omega)&= \lim_{i\Omega_n\rightarrow\omega+i0^+}I_{\M,\N}(\mathbf{q}, i\Omega_n) \nonumber \\
    &= \kappa_{\M\N} \cdot \frac{1}{N} \sum_{\k \in \text{MBZ}} \sum_{a,b=1} \Gamma(\k)_{\M a}^{*}\Gamma(\k)_{\N a}\Gamma(\mathbf{k+q})_{Mb}\Gamma(\mathbf{k+q})_{\N b}^{*} \left[\frac{n_{F}(\E_{a}(\k))\left[1-n_{F}(\E_{b}(\mathbf{k+q}))\right]}{\omega^{+}+(\E_{a}(\k)-\E_{b}(\mathbf{k+q}))}\right].    
\end{align}
Therefore, 
\begin{align}
    S({\bf q},\omega)=\sum_{\M,\N}I_{\M,\N}({\q},\omega)
\end{align}
as given in \Eq{eq:dyn_struct_fac} of the main text, and the corresponding plots are presented in \Fig{fig:Sq_plots} where we have set the temperature $T$ entering $n_F$ to an appropriately small value to account for contributions arising from only the occupied bands.
\fi

\end{document}